\documentclass[twocolumn,tighten]{aastex62}
\usepackage{amssymb, amsmath,framed}

\usepackage{bm}
\expandafter\ifx\csname package@font\endcsname\relax\else
 \expandafter\expandafter
 \expandafter\usepackage
 \expandafter\expandafter
 \expandafter{\csname package@font\endcsname}
\fi
\def\bq{\begin{equation}}
\def\eq{\end{equation}}
\def\bqy{\begin{eqnarray}}
\def\eqy{\end{eqnarray}}

\begin{document}
\title{\large{Searching the Moon for Extrasolar Material and the Building Blocks of Extraterrestrial Life}}

\correspondingauthor{Manasvi Lingam}
\email{manasvi.lingam@cfa.harvard.edu}

\author{Manasvi Lingam}
\affiliation{Institute for Theory and Computation, Harvard University, Cambridge MA 02138, USA}

\author{Abraham Loeb}
\affiliation{Institute for Theory and Computation, Harvard University, Cambridge MA 02138, USA}

\begin{abstract}
Due to its absence of an atmosphere and relative geological inertness, the Moon's surface records past impacts of objects from the Solar system and beyond. We examine the prospects for discovering extrasolar material near the lunar surface and predict that its abundance is $\mathcal{O}(10)$ parts-per-million (ppm). The abundances of extrasolar organic carbon and biomolecular building blocks (e.g., amino acids) are estimated to be on the order of $0.1$ ppm and $< 0.1$ parts-per-billion (ppb), respectively. We describe strategies for identifying extrasolar material and potentially detecting extrasolar biomolecular building blocks as well as molecular biosignatures of extinct extraterrestrial life. Thus, viewed collectively, we argue that \emph{in situ} lunar exploration can provide vital new clues for astrobiology. \\
\end{abstract}

\section{Introduction} \label{SecIntro}
The discovery of the interstellar object `Oumuamua in 2017 \citep{MWM} led to a substantial increase in the expected number density of interstellar objects relative to certain earlier estimates \citep{MTL09}. Recently, the identification of a putative interstellar meteor by \citet{SL19a} enabled the determination of the flux of extrasolar objects impacting the Earth's atmosphere \citep{SL19b}. 

There are several avenues to analyze objects which originate beyond the Solar system (in short, extrasolar). First, one can send out spacecrafts to investigate interstellar dust in the neighborhood of Earth \citep{LBG00}, unbound objects like `Oumuamua \citep{SL18}, gravitationally captured objects within our Solar system \citep{LL18}, or even nearby exoplanets like Proxima b.\footnote{\url{https://breakthroughinitiatives.org/initiative/3}} A second possibility entails remote sensing studies of interstellar meteors that burn up in Earth's atmosphere \citep{SL19b} or objects that graze the Sun \citep{FL19}. We will instead address a third route in this Letter: combing through lunar samples to search for extrasolar material. The same approach is utilizable, in principle, for detecting extrasolar material deposited on the surfaces of asteroids and comets.

It is very beneficial that extrasolar objects impact not only the Earth but also the Moon. The latter is advantageous from two different standpoints. First, as the Moon lacks an atmosphere, there is minimal ablation of small objects relative to Earth, consequently ensuring that they are preserved and do not burn up before impacting the surface. Second, it is well-known that the Moon is geologically inert with respect to the Earth over the past few Gyr \citep{JHA12}. This feature ensures that the Moon, unlike the Earth, preserves a comprehensive geological record dating back almost to its formation around $4.5$ Ga.

From a practical standpoint, the strategy of searching lunar samples has two benefits with respect to the alternatives mentioned earlier. First, the Apollo missions returned $\sim 400$ kg of lunar material to the Earth, ensuring that it is feasible to examine these samples for extrasolar debris. Second, both the federal and private sectors have expressed an interest in going back to the Moon in the upcoming decade,\footnote{\url{https://www.nasa.gov/specials/apollo50th/back.html}} and potentially establishing lunar bases in the long run.\footnote{\url{http://www.asi.org/}} There are numerous benefits expected to accrue from the sustained \emph{in situ} exploration of the Moon in areas as diverse as high-energy physics, medicine, planetary science and astrobiology \citep{Cock10,CAC12}. We suggest that one should also include the detection of extrasolar material - in particular, the search for the building blocks of extraterrestrial life - to the list of benefits from lunar exploration.

The outline of the Letter is as follows. We predict the mass and number flux of extrasolar impactors striking the lunar surface in Section \ref{SecMF}. We estimate the abundances of extrasolar material, organics, and biomolecular building blocks in Section \ref{SecAbE}. Next, we briefly outline methodologies by which the extrasolar components may be detected in Section \ref{SecSeaE}. Finally, we summarize our central results in Section \ref{SecConc}.

\section{Mass flux of Extrasolar Impactors}\label{SecMF}
Henceforth, we use the subscript `S' to reference impactors whose origin lies within the Solar system (i.e., intrasolar) and the subscript `E' to denote impactors that originate outside the solar system (i.e., extrasolar). 

We begin by assessing the number flux of extrasolar impactors on the Moon. In order to do so, we note that the contribution from gravitational focusing can be neglected since the correction factor $\left(1 + v_\mathrm{esc}^2/v_\infty^2\right)$ is close to unity, where $v_\mathrm{esc}$ is the escape velocity and $v_\infty$ represents the excess velocity at a large distance. The probability distribution function for the impact flux is denoted by $\mathcal{P}_E(m)$, in units of m$^{-2}$ s$^{-1}$ kg$^{-1}$, where $m$ is the mass of the impactor. We will work with a power-law function in the mass range of interest, i.e., $\mathcal{P}_E(m) = C_E m^{-\lambda_E}$, where $C_E$ is the proportionality constant and $\lambda_E$ is the power-law index. This ansatz allows us to determine the number flux of impactors $\dot{\mathcal{N}}_E(m)$ with masses $> m$ as follows:
\begin{equation}\label{PhiEI}
  \dot{\mathcal{N}}_E(m) = \int_m^\infty \mathcal{P}_E(m')\,dm' =   \frac{C_E}{|-\lambda_E + 1|} m^{-\lambda_E + 1}. 
\end{equation}
As we are interested in extraterrestrial impactors, we will make the assumption that the number flux is roughly the same for the Moon and the Earth's atmosphere. This is fairly valid because the extra contribution from the orbital velocity of the Moon is smaller than $v_\infty$ and $v_\mathrm{esc}$ by approximately an order of magnitude. Note that the total number of impacts per unit time is \emph{not} similar for both worlds, even if the number fluxes are comparable, because of their differing surface areas. In conjunction with the compiled data from Figure 1 and Section 2 of \citet{SL19b}, we estimate $\dot{\mathcal{N}}_E(m)$ to be
\begin{equation}\label{PhiEEmp}
  \dot{\mathcal{N}}_E(m) \sim 4.4 \times 10^{-22}\,\mathrm{m^{-2}\,s^{-1}}\, \left(\frac{m}{1\,\mathrm{kg}}\right)^{-1.14}.  
\end{equation}
As a consistency check, if we substitute $m = 10^{-14}$ kg in the above expression, we obtain $\dot{\mathcal{N}}_E \sim 4 \times 10^{-6}$ m$^{-2}$ s$^{-1}$. This result is in reasonable agreement with the empirical estimate of $\dot{\mathcal{N}}_E \sim 1 \times 10^{-6}$ m$^{-2}$ s$^{-1}$ based on \emph{in situ} measurements carried out by the Ulysses and Galileo spacecrafts \citep{LBG00}. In addition, the power-law exponent of $-1.14$ specified in (\ref{PhiEEmp}) exhibits very good agreement with the empirical value of $-1.1$ from spacecraft observations \citep{LBG00}. It is straightforward to determine $C_E$ and $\lambda_E$ from (\ref{PhiEI}) and (\ref{PhiEEmp}); for example, we find $\lambda_E = 2.14$.

In a similar fashion, we can determine the flux of Solar system objects that impact the Moon. We define $\mathcal{P}_S(m) = C_S m^{-\lambda_S}$ and thereby compute $\dot{\mathcal{N}}_S(m)$ in the same fashion as (\ref{PhiEI}). At very small masses, the values of $C_S$ and $\lambda_S$ are not tightly constrained. Older empirical measurements of interplanetary dust particles (IDPs) indicated that $\lambda_S \approx 2.34$ for $m > 10^{-7}$ kg \citep{GHS11}, whereas more recent studies based on the Lunar Dust Experiment (LDEX) onboard the Lunar Atmosphere and Dust Environment Explorer (LADEE) concluded that $\lambda_S \approx 1.9$ for dust grains with masses $> 10^{-15}$ kg \citep{SH16}. If we further suppose that the flux at Earth's atmosphere is comparable to that on the Moon, Figure 1 of \citet{BA06} indicates that $\lambda_S \approx 1.9$. Thus, we find that $\lambda_S$ is not very different from $\lambda_E$. We introduce the ansatz
\begin{equation}\label{PhiSEmp}
    \dot{\mathcal{N}}_S(m) \sim 6 \times 10^{-19}\,\mathrm{m^{-2}\,s^{-1}}\, \left(\frac{m}{1\,\mathrm{kg}}\right)^{-0.9},
\end{equation}
where the normalization constant is chosen to preserve consistency with Figure 1 of \citet{BA06}. For $m = 0.1$ kg, the above formula yields $\dot{\mathcal{N}}_S \sim 4.8 \times 10^{-18}$ m$^{-2}$ s$^{-1}$. By using the observational data in Figure 4 of \citet{GHS11}, we end up with $\dot{\mathcal{N}}_S \sim 6.3 \times 10^{-18}$ m$^{-2}$ s$^{-1}$, indicating that the above ansatz may be a reasonable estimate.

Along the same lines, we can determine the mass flux of impactors within a given mass range of $\left(m_\mathrm{min},\,m_\mathrm{max}\right)$. The corresponding mass flux, denoted by $\dot{\mathcal{M}}_{E,S}$, is
\begin{equation}
    \dot{\mathcal{M}}_{E,S} = \int_{m_\mathrm{min}}^{m_\mathrm{max}} m'\,\mathcal{P}_{E,S}(m')\,dm'.
\end{equation}
For our lower bound, we choose approximately $\mu$m-sized objects (with $m_\mathrm{min} = 10^{-15}$ kg) as they represent the smallest particles that may host organic material \citep{FKF03,Kwok}; in the most optimal circumstances, they might be capable of transporting living or extinct microbes \citep{Wes10}. Our upper bound of $m_\mathrm{max} = 10^{15}$ kg is based on the fact that objects with higher masses are unlikely to have impacted the Moon over its current age. The ratio of the two mass fluxes ($\delta_{ES}$) is defined as
\begin{equation}\label{RatFlux}
    \delta_{ES} \equiv \frac{\dot{\mathcal{M}}_E}{\dot{\mathcal{M}}_S} \sim 2.6 \times 10^{-3},
\end{equation}
where the last equality follows from employing the preceding relations. In other words, the mass flux of extrasolar objects striking the Moon is potentially three orders of magnitude smaller than the mass flux of impactors originating from within our Solar system. 

We caution that the scaling relations specified for $\dot{\mathcal{N}}_E$ and $\dot{\mathcal{N}}_S$ constitute merely heuristic estimates as they are subject to numerous uncertainties (most notably for the flux of extrasolar objects). It is likely that a single power-law function will not suffice, thereby necessitating the use of broken power-laws in future studies. Another simplification introduced herein is that the flux of impactors remains roughly constant over time. While this is approximately correct when it comes to intrasolar objects over the past few Gyr and possibly valid for extrasolar objects, it is \emph{not} valid for intrasolar objects during the early stages of our Solar system ($\gtrsim 4.0$ Ga), when the impact rates were a few orders of magnitudes higher \citep{CS92}.

\section{Abundance of Extrasolar Material on the Moon}\label{SecAbE}
The ratio $\delta_{ES}$ is valuable because it enables us to calculate the abundance of extrasolar material present near the lunar surface. However, in doing so, we rely upon the assumption that the gardening depths of intrasolar and extrasolar objects are comparable. This is not entirely unreasonable because the specific kinetic energy is proportional to $v_\mathrm{esc}^2 + v_\infty^2$, implying that its value for extrasolar objects is conceivably an order of magnitude higher than for intrasolar objects. 

If the variations in gardening depth are ignored, the abundance of extrasolar material by weight ($\phi_E$) is approximately proportional to $\dot{\mathcal{M}}_E$, consequently yielding $\phi_E \sim \delta_{ES} \phi_S$ with $\phi_S$ signifying the abundance of (micro)meteoritic material originating from the Solar system. Based on the analysis of lunar samples, it has been estimated that this component makes up $\sim 1$-$1.5\%$ (by weight) of the lunar soil and $\sim 1.28\%$ of the lunar regolith \citep{AGKM,MVT15}. Therefore, by using the above expression for $\phi_E$, we arrive at $\phi_E \sim 30$ ppm, namely, the mass fraction of extrasolar material is $\sim 3 \times 10^{-5}$. In comparison, material ejected from Earth subsequently deposited on the Moon is predicted to occur at an abundance of $\sim 1$-$2$ ppm at the surface \citep{Arm10}.

Of this extrasolar material, we note that $\sim 10^{-3}$, therefore amounting to an abundance of $\sim 30$ ppb, is derived from halo stars \citep{SL19c}. Another crucial point worth noting before proceeding further is that the preservation of older extrasolar material is feasible in principle because the Moon has been geologically inactive relative to Earth during the past few Gyr \citep{JHA12}. If we suppose, for instance, that the material is uniformly distributed over time and adequately preserved, we find that $\sim 10\%$ of all extrasolar material would have been deposited $> 4$ Ga. In other words, the abundance of such material might be $\sim 3$ ppm after using the previous result for $\phi_E$.

However, the extrasolar material deposited on the surface will comprise both inorganic and organic components. It is very difficult to estimate the abundance of the latter as we lack precise constraints on the abundance of organics in ejecta expelled from extrasolar systems as well as the likelihood of their survival during transit and impact with the lunar surface. Hence, our subsequent discussion must be viewed with due caution as we operate under the premise that (micro)meteorites and IDPs within the Solar system are not very atypical relative to other planetary systems.\footnote{This line of reasoning goes by many names, including the Copernican Principle and the Principle of Mediocrity, and is often implicitly invoked in astrobiology.}

We begin by considering the abundance of extrasolar organic material. Even within the Solar system, the inventory of organic carbon varies widely across meteorites and IDPs. For instance, it is believed that organic carbon comprises $\sim 1.5$-$4\%$ by weight in carbonaceous chondrites \citep{PS10}, whereas it is lower for other classes of meteorites. When it comes to IDPs, laboratory analyses indicate that they possess $\sim 10\%$ carbon by weight on average \citep{CS92,PCF06}. If we err on the side of caution and choose a mean value of $\sim 1\%$ as not all carbon is incorporated in organic material, we find that the abundance of extrasolar organic material ($\phi_{E,O}$) may be $\sim 3 \times 10^{-7}$, namely, we obtain $\phi_{E,O} \sim 0.3$ ppm.

However, it should be noted that the majority of organic carbon ($> 70\%$) in carbonaceous chondrites is locked up in the form of insoluble compounds that are ``kerogen-like'' in nature \citep{PS10,QOB14}. As organics constitute a very broad category, it is more instructive to focus on specific classes. We will henceforth mostly restrict ourselves to amino acids because they are building blocks for proteins and are therefore essential for life-as-we-know-it. Other organic compounds that were identified in meteorites include aliphatic and aromatic hydrocarbons, phosphonic and sulfonic acids, and polyols. 

We begin by considering the abundance of amino acids. Meteorites exhibit different concentrations of amino acids with values ranging from $\ll 1$ ppm to $\gtrsim 100$ ppm \citep{MAO}. The uncertainty for IDPs is even larger because only a few amino acids such as $\alpha$-amino isobutyric acid have been detected and the average abundance of amino acids in IDPs remains poorly constrained \citep{MPT04}. Hence, we will resort to an alternative strategy instead. The analysis of lunar samples from the Apollo missions indicates that the concentration of amino acids is $\sim 0.1$-$100$ ppb with typical values on the order of $\sim 10$ ppb \citep{GZK72,HHW71,ECD16}. 

Earlier, we determined that the extrasolar mass flux is lower by three orders of magnitude compared to the intrasolar mass flux. Hence, using the value of $\delta_{ES}$ from (\ref{RatFlux}), we find that the concentration of extrasolar amino acids is potentially $\sim 30$ parts-per-trillion (ppt). However, this estimate is an upper bound in all likelihood because it presumes that the fiducial choice of $10$ ppb for amino acids in the lunar regolith arises solely from (micro)meteorite impacts. In actuality, on account of the high enantiomeric excesses detected, it is believed these samples have experienced some terrestrial biological contamination \citep{ECD16}. 

In analogy with the discovery of carboxylic acids and nucleobases - the building blocks of lipids and nucleic acids, respectively - in meteorites on Earth, it is plausible that these compounds might be found on the Moon. For example, analysis of meteorites has revealed that carboxylic acids may comprise $\sim 40$-$300$ ppm \citep{PCF06}. Adopting a fiducial value of $\sim 10$ ppm for carboxylic acids in extrasolar material by erring on the side of caution, we estimate an abundance of $\sim 0.3$ ppb for extrasolar carboxylic acids in the lunar regolith after using the prior estimate for $\phi_E$. A similar analysis can be carried out for nucleobases by employing carbonaceous chondrites as a proxy. Choosing a nucleobase abundance of $\sim 0.1$ ppm in chondrites \citep{CSC11}, we obtain an estimate of $\sim 3$ ppt for extrasolar nucleobases near the lunar surface. 

We reiterate that the numbers described herein are rough estimates because a number of key processes are not tightly constrained. Apart from the direct contribution of extrasolar objects impacting the Moon, it is possible for extrasolar material to be deposited on intrasolar objects that subsequently impact the Moon and thereby deposit this material on the lunar surface. It is likely, however, that this contribution will be sub-dominant.

\section{Searching for Extrasolar Material on the Moon}\label{SecSeaE}
Hitherto, we have calculated the abundance of extrasolar material deposited on the lunar surface. However, this raises an immediate question: how do we distinguish between material (e.g., micrometeorites and IDPs) derived from within and outside the Solar system?

The solution may lie, at least partly, in analyzing multiple isotope ratios of samples \citep{LL18}. Of the various candidates, perhaps the best studied are the oxygen isotope ratios. In the oxygen three-isotope plot, involving the isotope ratios $^{17}$O/$^{16}$O and $^{18}$O/$^{16}$O, the terrestrial fractionation line has a slope of approximately $0.5$ whereas carbonaceous chondrites are characterized by a slope of $\sim 1$
\citep{Clay03,KA09}. It should also be noted that the $^{17}$O/$^{18}$O ratio exhibits a lower value in the Solar system in comparison to the Galactic average \citep{NG12}. Thus, significant deviations from the Solar system values in the oxygen three-isotope plot might imply that the sample is extrasolar in origin. 

Apart from oxygen isotopes, other extrasolar flags include carbon and nitrogen isotope ratios, corresponding to $^{12}$C/$^{13}$C and $^{14}$N/$^{15}$N, respectively \citep{Mum11,FM15}. Note, for instance, that enhanced values of the $^{12}$C/$^{13}$C ratio could arise in extrasolar objects that have traversed through regions in proximity to Young Stellar Objects \citep{SPY15}. In addition to isotope ratios, anomalies in CN-to-OH ratios as well as the abundances of bulk elements, C$_2$ and C$_3$ molecules might also serve as effective methods for discerning extrasolar material \citep{LS07,Sch08}. 

Once the identification of extrasolar grains has been achieved, one could attempt to identify the organics present within them. A plethora of standard techniques can be employed such as liquid chromatography-mass spectrometry. Using such procedures, the identification of amino acids, nucleobases and other organic compounds is feasible at sub-ppb concentrations \citep{GDA06,CSC11,BSE12}. The detection of either nucleobases or amino acids that are neither prevalent in terrestrial nor meteoritic material would lend further credence to the notion that the sample under question may be extrasolar in nature.\footnote{It is worth appreciating that meteorites contain ``exotic'' organic compounds that are very rare on Earth. For instance, the analysis of carbonaceous meteorites has revealed the existence of nucleobase analogs (e.g., purine) whose abundances are extremely low on Earth \citep{CSC11}.}

Hitherto, we have limited our discussion to extrasolar material and organic compounds. There is yet another scenario worth mentioning, albeit with a potentially much lower probability, namely, the detection of biosignatures corresponding to extinct extraterrestrial life.\footnote{We have implicitly excluded the prospects for living extraterrestrial organisms because the Moon's habitability ``window'' appears to have come to a close just millions of years after its formation \citep{SMC18}.} There are a number of methods that may be utilized to search for biomarkers. Some of the measurable characteristics of molecular biosignatures include: (a) enantiomeric excesses stemming from homochirality, (b) preference for certain diastereoisomers and structural isomers, and (c) isotopic heterogeneities at molecular or sub-molecular levels \citep{SAM08}.  A review of numerous life-detection experiments and their efficacy can be found in \citet{NHV18}. The most ideal scenario arguably entails the discovery of extrasolar microfossils as they would provide clear-cut evidence for extraterrestrial life; on Earth, the oldest microfossils with unambiguous evidence of cell lumens and walls are from the $\sim 3.4$ Ga Strelley Pool Formation in Western Australia \citep{WKS11}.

\section{Conclusion}\label{SecConc}
In light of recent discoveries of interstellar objects, we have studied the deposition of extrasolar material on the lunar surface by estimating the mass fluxes of impactors originating from within and outside our Solar system. Our choice of the Moon is motivated by the fact that it lacks an atmosphere (avoiding ablation of the impactors) and is mostly geologically inactive (allowing for long-lived retention of material). 

Our calculations suggest that the abundance of extrasolar material at the surface is $\sim 30$ ppm, with the abundance of detritus deposited $> 4$ Ga being $\sim 3$ ppm. Of this material, a small fraction will exist in the form of organic molecules. We estimated that the abundance of extrasolar organic carbon near the lunar surface is $\sim 0.3$ ppm. Among the various organic compounds, the abundances of carboxylic acids, amino acids and nucleobases are of particular interest as they constitute the building blocks for life-as-we-know-it. Our results indicate that their maximal abundances might be $\sim 300$ ppt, $\sim 30$ ppt and $\sim 3$ ppt, respectively.

We outlined how the detection of extrasolar debris may be feasible by analyzing lunar samples. A combination of isotope ratios (oxygen in particular), elemental abundances, and other diagnostics might allow us to identify extrasolar material on the Moon. This material can then be subjected to subsequent laboratory experiments to search for organic compounds such as amino acids as well as molecular biosignatures arising from extinct extraterrestrial life. Altogether, these analyses could provide important new clues for astrobiology.

Even the ``mere'' discovery of inorganic extrasolar material will open up new avenues for research. In particular, by studying the chemical composition of this material, it may be possible to place constraints on planetary formation models, assess the habitability of early planetary systems, gauge the origin and evolution of exo-Oort clouds, and determine the chemical diversity of extrasolar planetary systems. Hence, a new channel for understanding these physical processes, separate from studying unbound interstellar objects such as `Oumuamua \citep{TRR17,RAV18,MM19}, can be initiated.

The discovery of extrasolar organics could reveal new complex macromolecules that may possess practical value in medicine and engineering. Furthermore, the detection of such molecules would enable us to gain a deeper understanding of what types of organics were synthesized in other planetary systems, allowing us to gauge the latter's prospects for hosting life. Finally, the discovery of molecular biosignatures confirming the existence of (extinct) extraterrestrial life will indubitably have far-reaching consequences for humankind. In view of these potential benefits, we contend that there are additional compelling grounds for sustained \emph{in situ} exploration of the lunar surface in the upcoming decades.

\acknowledgments
This work was supported in part by the Breakthrough Prize Foundation, Harvard University's Faculty of Arts and Sciences, and the Institute for Theory and Computation (ITC) at Harvard University.


\begin{thebibliography}{}
\expandafter\ifx\csname natexlab\endcsname\relax\def\natexlab#1{#1}\fi
\providecommand{\url}[1]{\href{#1}{#1}}

\bibitem[{{Anders} {et~al.}(1973){Anders}, {Ganapathy}, {Kr{\"a}henb{\"u}hl},
  \& {Morgan}}]{AGKM}
{Anders}, E., {Ganapathy}, R., {Kr{\"a}henb{\"u}hl}, U., \& {Morgan}, J.~W.
  1973, Moon, 8, 3

\bibitem[{{Armstrong}(2010)}]{Arm10}
{Armstrong}, J.~C. 2010, Earth Moon Planets, 107, 43

\bibitem[{{Bland} \& {Artemieva}(2006)}]{BA06}
{Bland}, P.~A., \& {Artemieva}, N.~A. 2006, Meteorit. Planet. Sci., 41, 607

\bibitem[{{Burton} {et~al.}(2012){Burton}, {Stern}, {Elsila}, {Glavin}, \&
  {Dworkin}}]{BSE12}
{Burton}, A.~S., {Stern}, J.~C., {Elsila}, J.~E., {Glavin}, D.~P., \&
  {Dworkin}, J.~P. 2012, Chem. Soc. Rev., 41, 5459

\bibitem[{{Callahan} {et~al.}(2011){Callahan}, {Smith}, {Cleaves}, {Ruzicka},
  {Stern}, {Glavin}, {House}, \& {Dworkin}}]{CSC11}
{Callahan}, M.~P., {Smith}, K.~E., {Cleaves}, H.~J., {et~al.} 2011, Proc. Natl.
  Acad. Sci. USA, 108, 13995

\bibitem[{{Chyba} \& {Sagan}(1992)}]{CS92}
{Chyba}, C., \& {Sagan}, C. 1992, Nature, 355, 125

\bibitem[{{Clayton}(2003)}]{Clay03}
{Clayton}, R.~N. 2003, Space Sci. Rev., 106, 19

\bibitem[{{Cockell}(2010)}]{Cock10}
{Cockell}, C.~S. 2010, Earth Moon Planets, 107, 3

\bibitem[{{Crawford} {et~al.}(2012){Crawford}, {Anand}, {Cockell}, {Falcke},
  {Green}, {Jaumann}, \& {Wieczorek}}]{CAC12}
{Crawford}, I.~A., {Anand}, M., {Cockell}, C.~S., {et~al.} 2012, Planet. Space
  Sci., 74, 3

\bibitem[{{Elsila} {et~al.}(2016){Elsila}, {Callahan}, {Dworkin}, {Glavin},
  {McLain}, {Noble}, \& {Gibson}}]{ECD16}
{Elsila}, J.~E., {Callahan}, M.~P., {Dworkin}, J.~P., {et~al.} 2016, Geochim.
  Cosmochim. Acta, 172, 357

\bibitem[{{Flynn} {et~al.}(2003){Flynn}, {Keller}, {Feser}, {Wirick}, \&
  {Jacobsen}}]{FKF03}
{Flynn}, G.~J., {Keller}, L.~P., {Feser}, M., {Wirick}, S., \& {Jacobsen}, C.
  2003, Geochim. Cosmochim. Acta, 67, 4791

\bibitem[{{Forbes} \& {Loeb}(2019)}]{FL19}
{Forbes}, J.~C., \& {Loeb}, A. 2019, Astrophys. J. Lett., 875, L23

\bibitem[{{F{\"u}ri} \& {Marty}(2015)}]{FM15}
{F{\"u}ri}, E., \& {Marty}, B. 2015, Nat. Geosci., 8, 515

\bibitem[{{Gehrke} {et~al.}(1972){Gehrke}, {Zumwalt}, {Kuo}, {Rash}, {Aue},
  {Stalling}, {Kenvolden}, \& {Ponnamperuma}}]{GZK72}
{Gehrke}, C.~W., {Zumwalt}, R.~W., {Kuo}, K., {et~al.} 1972, Space Life Sci.,
  3, 439

\bibitem[{{Glavin} {et~al.}(2006){Glavin}, {Dworkin}, {Aubrey}, {Botta},
  {Doty}, {Martins}, \& {Bada}}]{GDA06}
{Glavin}, D.~P., {Dworkin}, J.~P., {Aubrey}, A., {et~al.} 2006, Meteorit.
  Planet. Sci., 41, 889

\bibitem[{{Gr{\"u}n} {et~al.}(2011){Gr{\"u}n}, {Horanyi}, \&
  {Sternovsky}}]{GHS11}
{Gr{\"u}n}, E., {Horanyi}, M., \& {Sternovsky}, Z. 2011, Planet. Space Sci.,
  59, 1672

\bibitem[{{Harada} {et~al.}(1971){Harada}, {Hare}, {Windsor}, \& {Fox}}]{HHW71}
{Harada}, K., {Hare}, P.~E., {Windsor}, C.~R., \& {Fox}, S.~W. 1971, Science,
  173, 433

\bibitem[{{Jaumann} {et~al.}(2012){Jaumann}, {Hiesinger}, {Anand}, {Crawford},
  {Wagner}, {Sohl}, {Jolliff}, {Scholten}, {Knapmeyer}, {Hoffmann}, {Hussmann},
  {Grott}, {Hempel}, {K{\"o}hler}, {Krohn}, {Schmitz}, {Carpenter},
  {Wieczorek}, {Spohn}, {Robinson}, \& {Oberst}}]{JHA12}
{Jaumann}, R., {Hiesinger}, H., {Anand}, M., {et~al.} 2012, Planet. Space Sci.,
  74, 15

\bibitem[{{Krot} {et~al.}(2009){Krot}, {Amelin}, {Bland}, {Ciesla}, {Connelly},
  {Davis}, {Huss}, {Hutcheon}, {Makide}, {Nagashima}, {Nyquist}, {Russell},
  {Scott}, {Thrane}, {Yurimoto}, \& {Yin}}]{KA09}
{Krot}, A.~N., {Amelin}, Y., {Bland}, P., {et~al.} 2009, Geochim. Cosmochim.
  Acta, 73, 4963

\bibitem[{{Kwok}(2019)}]{Kwok}
{Kwok}, S. 2019, Res. Astron. Astrophys., 19, 049

\bibitem[{{Landgraf} {et~al.}(2000){Landgraf}, {Baggaley}, {Gr{\"u}n},
  {Kr{\"u}ger}, \& {Linkert}}]{LBG00}
{Landgraf}, M., {Baggaley}, W.~J., {Gr{\"u}n}, E., {Kr{\"u}ger}, H., \&
  {Linkert}, G. 2000, J. Geophys. Res., 105, 10343

\bibitem[{{Langland-Shula} \& {Smith}(2007)}]{LS07}
{Langland-Shula}, L.~E., \& {Smith}, G.~H. 2007, Astrophys. J. Lett., 664, L119

\bibitem[{{Lingam} \& {Loeb}(2018)}]{LL18}
{Lingam}, M., \& {Loeb}, A. 2018, Astron. J., 156, 193

\bibitem[{{Martins} {et~al.}(2007){Martins}, {Alexander}, {Orzechowska},
  {Fogel}, \& {Ehrenfreund}}]{MAO}
{Martins}, Z., {Alexander}, C.~M.~O.~D., {Orzechowska}, G.~E., {Fogel}, M.~L.,
  \& {Ehrenfreund}, P. 2007, Meteorit. Planet. Sci., 42, 2125

\bibitem[{{Matrajt} {et~al.}(2004){Matrajt}, {Pizzarello}, {Taylor}, \&
  {Brownlee}}]{MPT04}
{Matrajt}, G., {Pizzarello}, S., {Taylor}, S., \& {Brownlee}, D. 2004,
  Meteorit. Planet. Sci., 39, 1849

\bibitem[{{McCubbin} {et~al.}(2015){McCubbin}, {Vander Kaaden}, {Tart{\`e}se},
  {Klima}, {Liu}, {Mortimer}, {Barnes}, {Shearer}, {Treiman}, {Lawrence},
  {Elardo}, {Hurley}, {Boyce}, \& {Anand}}]{MVT15}
{McCubbin}, F.~M., {Vander Kaaden}, K.~E., {Tart{\`e}se}, R., {et~al.} 2015,
  Am. Mineral., 100, 1668

\bibitem[{{Meech} {et~al.}(2017){Meech}, {Weryk}, {Micheli}, {Kleyna},
  {Hainaut}, {Jedicke}, {Wainscoat}, {Chambers}, {Keane}, {Petric}, {Denneau},
  {Magnier}, {Berger}, {Huber}, {Flewelling}, {Waters}, {Schunova-Lilly}, \&
  {Chastel}}]{MWM}
{Meech}, K.~J., {Weryk}, R., {Micheli}, M., {et~al.} 2017, Nature, 552, 378

\bibitem[{{Moro-Mart{\'{\i}}n}(2019)}]{MM19}
{Moro-Mart{\'{\i}}n}, A. 2019, Astron. J., 157, 86

\bibitem[{{Moro-Mart{\'{\i}}n} {et~al.}(2009){Moro-Mart{\'{\i}}n}, {Turner}, \&
  {Loeb}}]{MTL09}
{Moro-Mart{\'{\i}}n}, A., {Turner}, E.~L., \& {Loeb}, A. 2009, Astrophys. J.,
  704, 733

\bibitem[{{Mumma} \& {Charnley}(2011)}]{Mum11}
{Mumma}, M.~J., \& {Charnley}, S.~B. 2011, Annu. Rev. Astron. Astrophys., 49,
  471

\bibitem[{{Neveu} {et~al.}(2018){Neveu}, {Hays}, {Voytek}, {New}, \&
  {Schulte}}]{NHV18}
{Neveu}, M., {Hays}, L.~E., {Voytek}, M.~A., {New}, M.~H., \& {Schulte}, M.~D.
  2018, Astrobiology, 18, 1375

\bibitem[{{Nittler} \& {Gaidos}(2012)}]{NG12}
{Nittler}, L.~R., \& {Gaidos}, E. 2012, Meteorit. Planet. Sci, 47, 2031

\bibitem[{{Pizzarello} {et~al.}(2006){Pizzarello}, {Cooper}, \&
  {Flynn}}]{PCF06}
{Pizzarello}, S., {Cooper}, G.~W., \& {Flynn}, G.~J. 2006, {The Nature and
  Distribution of the Organic Material in Carbonaceous Chondrites and
  Interplanetary Dust Particles}, ed. D.~S. {Lauretta} \& H.~Y. {McSween} (The
  University of Arizona Press), 625--651

\bibitem[{{Pizzarello} \& {Shock}(2010)}]{PS10}
{Pizzarello}, S., \& {Shock}, E. 2010, Cold Spring Harb. Perspect. Biol., 2,
  a002105

\bibitem[{{Quirico} {et~al.}(2014){Quirico}, {Orthous-Daunay}, {Beck}, {Bonal},
  {Brunetto}, {Dartois}, {Pino}, {Montagnac}, {Rouzaud}, {Engrand}, \&
  {Duprat}}]{QOB14}
{Quirico}, E., {Orthous-Daunay}, F.-R., {Beck}, P., {et~al.} 2014, Geochim.
  Cosmochim. Acta, 136, 80

\bibitem[{{Raymond} {et~al.}(2018){Raymond}, {Armitage}, {Veras}, {Quintana},
  \& {Barclay}}]{RAV18}
{Raymond}, S.~N., {Armitage}, P.~J., {Veras}, D., {Quintana}, E.~V., \&
  {Barclay}, T. 2018, Mon. Not. R. Astron. Soc., 476, 3031

\bibitem[{{Schleicher}(2008)}]{Sch08}
{Schleicher}, D.~G. 2008, Astron. J., 136, 2204

\bibitem[{{Schulze-Makuch} \& {Crawford}(2018)}]{SMC18}
{Schulze-Makuch}, D., \& {Crawford}, I.~A. 2018, Astrobiology, 18, 985

\bibitem[{{Seligman} \& {Laughlin}(2018)}]{SL18}
{Seligman}, D., \& {Laughlin}, G. 2018, Astron. J., 155, 217

\bibitem[{{Siraj} \& {Loeb}(2019{\natexlab{a}})}]{SL19a}
{Siraj}, A., \& {Loeb}, A. 2019{\natexlab{a}}, submitted to Astrophys. J.
  Lett., arXiv:1904.07224

\bibitem[{{Siraj} \& {Loeb}(2019{\natexlab{b}})}]{SL19b}
---. 2019{\natexlab{b}}, submitted to Astrophys. J. Lett., arXiv:1906.03270

\bibitem[{{Siraj} \& {Loeb}(2019{\natexlab{c}})}]{SL19c}
---. 2019{\natexlab{c}}, submitted to Mon. Not. R. Astron. Soc. Lett.,
  arXiv:1906.05291

\bibitem[{{Smith} {et~al.}(2015){Smith}, {Pontoppidan}, {Young}, \&
  {Morris}}]{SPY15}
{Smith}, R.~L., {Pontoppidan}, K.~M., {Young}, E.~D., \& {Morris}, M.~R. 2015,
  Astrophys. J., 813, 120

\bibitem[{{Summons} {et~al.}(2008){Summons}, {Albrecht}, {McDonald}, \&
  {Moldowan}}]{SAM08}
{Summons}, R.~E., {Albrecht}, P., {McDonald}, G., \& {Moldowan}, J.~M. 2008,
  Space Sci. Rev., 135, 133

\bibitem[{{Szalay} \& {Hor{\'a}nyi}(2016)}]{SH16}
{Szalay}, J.~R., \& {Hor{\'a}nyi}, M. 2016, Geophys. Res. Lett., 43, 4893

\bibitem[{{Trilling} {et~al.}(2017){Trilling}, {Robinson}, {Roegge},
  {Chandler}, {Smith}, {Loeffler}, {Trujillo}, {Navarro-Meza}, \&
  {Glaspie}}]{TRR17}
{Trilling}, D.~E., {Robinson}, T., {Roegge}, A., {et~al.} 2017, Astrophys. J.
  Lett., 850, L38

\bibitem[{{Wacey} {et~al.}(2011){Wacey}, {Kilburn}, {Saunders}, {Cliff}, \&
  {Brasier}}]{WKS11}
{Wacey}, D., {Kilburn}, M.~R., {Saunders}, M., {Cliff}, J., \& {Brasier}, M.~D.
  2011, Nat. Geosci., 4, 698

\bibitem[{{Wesson}(2010)}]{Wes10}
{Wesson}, P.~S. 2010, Space Sci. Rev., 156, 239

\end{thebibliography}

\end{document}